\documentclass[12pt]{iopart}
\usepackage{dcolumn}
\usepackage{bm}
\newcommand{\lsm}{L$\sigma$M}

\newcommand{\be}{\begin{equation}}
\newcommand{\ee}{\end{equation}}
\newcommand{\bea}{\begin{eqnarray}}
\newcommand{\eea}{\end{eqnarray}}

\newcommand{\refc}[1]{Ref.~\cite{#1}}

\newcommand{\prp}{{\it Preprint}}
\begin{document}
\title[Constituent Quark Masses and the Electroweak Standard Model]
{Constituent Quark Masses and  \\ the Electroweak Standard Model}
\author{M.\ D.\ Scadron}
\address{Physics Department, University of Arizona, Tucson, AZ 85721, USA} 
\ead{scadron@physics.arizona.edu}
\author{R.\ Delbourgo}
\address{School of Mathematics and Physics, University of Tasmania \\
GPO Box 252-21, Hobart 7001, Australia}
\ead{Bob.Delbourgo@utas.edu.au}
\author{G.\ Rupp}
\address{Centro de F\'{\i}sica das Interac\c{c}\~{o}es Fundamentais,
Instituto Superior T\'{e}cnico, P-1049-001 Lisboa, Portugal} 
\ead{george@ist.utl.pt}
\date{\today}

\begin{abstract}

Constituent quark masses can be determined quite well from experimental data
in several ways and one can obtain fairly accurate values for all six $m_q$.
The strong quark-meson coupling $g=2\pi /\sqrt{3}$ arises from the quark-level
linear $\sigma$ model, whereas $e$ and $\sin \theta_w$ arise from weak
interactions when the heavy $M_W$ and $M_Z$ are regarded as resonances in
analogy with the strong KSFR relation. The Higgs boson mass, tied to null 
expectation value of charged Higgs components, is found to be around 317 GeV.
Finally, the experimental CPV phase angle $\delta$ and the three CKM angles 
$\Theta_c, \Theta_2, \Theta_3$ are successfully deduced from the 6 constituent
quark masses following Fritzsch's approach.

\end{abstract}

\maketitle

\section{Introduction}

One of the principal complaints about the electroweak standard model (EWSM) is 
that it contains too many parameters, in fact no fewer than 19 of them, even 
if one disregards massive neutrinos. Aside from this plethora, there is the
matter of the quark masses being ``current'' masses, and somewhat far removed
from effective ``constituent'' masses due to dynamical QCD contributions, 
which may in principle change with quark flavour, because the current masses are
nonzero. The values of current quark masses are often the subject of debate,
but much less controversy surrounds the constituent quarks. The purpose
of this paper is to demonstrate the usefulness of naive constituents to
shed light on and reduce some of the parameters of the standard model.

In Secs.~2--5 we show that the $u,d,s,c,b$ constituent masses may be reliably 
found by various experiments and are perfectly consistent with
chiral Goldberger-Treiman relations (GTRs). In Sec.~6 we indicate the 
analogies between strong interactions and the EWSM, by regarding the $Z$ and
$W$ bosons as resonances, when invoking strong-interaction VMD and KSRF-type
relations. In our treatment the $W$ couples to sources in ($V\!-\!A$) form 
(as originally found for low-mass hadrons and leptons by Sudarshan \& Marshak 
and Feynman \& Gell-Mann), while $Z$ coupling to
leptons is largely axial. As a by-product we predict $m_t\approx176.7$ GeV, 
via a GTR construction, and arrive at a plausible Higgs mass $M_H\approx316.7$ 
GeV. Finally, in Sec.~7 we attempt to ``predict'' the mixing angles, by 
following Fritzsch's approach, but using constituent quarks instead of current 
quarks, and show that the agreement with experiment is quite reasonable. 

\section{Light-quark mass difference}
The simplest way \cite{DLS99} to estimate the $u$-$d$ quark mass difference 
is via the neutral- and charged-kaon mass difference (neglecting the small 
error \cite{PDG04}):
\be
\fl m_d-m_u \approx m_{\bar{s}d}-m_{\bar{s}u} = m_{K^0}-m_{K^+}=  
(497.648-493.677) \:\mbox{MeV}\approx3.97\:\mbox{MeV}  .
\label{kaondu}
\ee
The kaon mass itself follows from knowledge of the chiral-breaking current
masses $m_n^{\mbox{\scriptsize cur}}$ ($n=u,d$) and
$m_s^{\mbox{\scriptsize cur}}$, but for the kaon mass difference we need
predict only the constituent-quark mass difference $m_d-m_u$.
Equation~(\ref{kaondu}) is compatible with the charged-$\Sigma$ baryon mass
difference
\be
\fl 2(m_d-m_u)\approx m_{dds}-m_{uus}=m_{\Sigma^-}\!-m_{\Sigma^+}=
(1197.45-1189.37)\:\mbox{MeV}\approx8.08\:\mbox{MeV} ,
\label{tsigmadu}
\ee
or
\be
m_d-m_u\approx4.04\:\mbox{MeV} \; .
\label{sigmadu}
\ee
If we include the $\Sigma^0$ in the latter estimate, we get the mass splittings
\cite{PDG04} 
$m_{\Sigma^-}(dds)-m_{\Sigma^0}(uds)=4.81$ MeV and
$m_{\Sigma^0}(uds)-m_{\Sigma^+}(uus)=3.27$ MeV, with average
constituent-quark mass difference $m_d-m_u=4.04$ MeV, which is equal to
(\ref{sigmadu}) and close to (\ref{kaondu}).

Besides the 4 MeV mass scale from (\ref{kaondu}) and (\ref{sigmadu}), an
approximate $m_d-m_u$ mass difference follows from higher resonances, namely
\be
\begin{array}{lcc}
\fl m_d-m_u\approx m_{\bar{s}d}-m_{\bar{s}u}=m_{K^{*0}}-m_{K^{*+}} &=&
(896.10-891.66) \:\mbox{MeV}\approx4.44\:\mbox{MeV} , \\ 
\fl m_d-m_u\approx m_{\bar{d}c}-m_{\bar{u}c}=m_{D^+}-m_{D^0} &=&
(1869.3-1864.5) \:\mbox{MeV}\approx4.8\:\mbox{MeV}  ,
\end{array}
\label{ksd}
\ee
though it is noticeable that these estimates deteriorate as the masses rise.
[Parenthetically, it is reassuring that (1) is in agreement with the nucleon 
mass difference; this can be estimated from the Coleman-Glashow \cite{CG64} 
$\lambda^3$ tadpole for $n$-$p$ nucleons
\cite{DLS99}
\be
\left(H^3_{\mbox{\scriptsize tad}}\right)_{n-p} \; \approx \;
2.5\;\mbox{MeV} \; ,
\label{tadpole}
\ee
less the proton current-current Hamiltonian density
\be
(H_{JJ})_p \; = \; \frac{3}{2}\frac{\alpha}{\pi}m_p\left[
\ln(\frac{\Lambda}{m_p})+\frac{1}{4}\right] \; \approx \;
1.2\;\mbox{MeV} \; ,
\label{hjj}
\ee
for UV cutoff $\Lambda\approx1.05$ GeV. Then
\be
m_n - m_p \; \approx \; 2.5\;\mbox{MeV}-1.2\;\mbox{MeV} \; = \;
1.3\;\mbox{MeV} \; ,
\label{mnp}
\ee
very near data \cite{PDG04} $m_n-m_p\approx1.293$ MeV.
Since (\ref{hjj}) requires a cutoff, we note that \cite{DLS99} predicts
a similar UV cutoff $\Lambda=1.02$ GeV for (squared) pion masses.]

Or we can justify the usual $\Delta I=1$ group-theory predictions in
nuclear physics to support Ref.~\cite{DLS99}, via a constituent-quark loop,
for $m_d-m_u=4$ MeV \cite{CS00}.

\section{Light-quark mass sum}
One way to get at the average nonstrange quark mass $\hat{m}\equiv(m_u+m_d)/2$ 
is to employ the nonrelativistic quark model (NRQM) \cite{BLP64} connecting
nucleon magnetic moments with those of the underlying quarks:
\be
\mu_p=\frac{4}{3}\,\mu_u-\frac{1}{3}\,\mu_d \;\;\; , \;\;\; 
\mu_n=\frac{4}{3}\,\mu_d-\frac{1}{3}\,\mu_u \;\;\; , 
\label{mupn}
\ee
where
$\mu_i=e_i/2m_i$ for constituent quark masses, with quark charges
$e_u=2e/3$, $e_d=-e/3$, and nucleon Bohr magneton scaled to $e/2M_N$.
Equations~(\ref{mupn}) tell us that
\be
\mu_p=\frac{e}{18}\left[\frac{8}{m_u}+\frac{1}{m_d}\right] \;\;\; ,
\;\;\; \mu_n=-\frac{e}{18}\left[\frac{4}{m_d}+\frac{2}{m_u}\right]\;.
\label{mpmn}
\ee
Since $m_d-m_u\approx4$ MeV (see Sec.~2), from $\mu_p$ in (\ref{mpmn}) 
and the experimental \cite{PDG04} value $\mu_p=2.792847\,e/2m_p$, we obtain 
(in MeV) a quadratic equation for $\hat{m}$: 
\be
\hat{m}^2-\frac{m_p}{2.792847}\:(\hat{m}+\frac{14}{9}) \; = \; 0 \; ,
\label{mhs}
\ee
whose only positive solution reads
\be
\hat{m} \; \approx \; 337.5\:\mbox{MeV} \; .
\label{mh}
\ee
Thus the light quark masses are
\be
m_d \; \approx \; 339.5\:\mbox{MeV} \;\;\; , \;\;\;
m_u \; \approx \; 335.5\:\mbox{MeV} \; .
\label{mud}
\ee
Backtracking to (\ref{mpmn}), the ratio
\be
-\frac{\mu_n}{\mu_p} \; = \; \frac{4m_u+2m_d}{m_u+8m_d} \; \approx \; 0.6623 
\label{mmnmpt}
\ee
is close to data \cite{PDG04}
\be
-\frac{\mu_n}{\mu_p} \; = \; \frac{1.913043}{2.792847} \; \approx \; 0.6850 \; .
\label{mmnmpe}
\ee

A second way of obtaining $\hat{m}$ is through the chiral-symmetric GTR
\be
\hat{m} \; \approx \; f_\pi g \; \approx \; (93\:\mbox{MeV})\,
\frac{2\pi}{\sqrt{3}} \; \approx \; 337.4\;\mbox{MeV} \; ,
\label{gtr}
\ee
which is satisfyingly close to (\ref{mh}). Here, $f_\pi\approx93$ MeV, and the
meson-quark coupling is $g=2\pi/\sqrt{3}$, determined from infrared QCD
\cite{ES84}, the $Z=0$ compositeness condition \cite{S62}, or the
quark-level linear $\sigma$ model (QLL$\sigma$M) \cite{DS95,SKRB03}.

\section{Strange quark mass}
Because the nonstrange GTR (\ref{gtr}) matches the magnetic-moment prediction
(\ref{mh}) so nicely, it invites us to extend the GTR to kaons, i.e.,
\be
\frac{1}{2}\,(m_s+\hat{m}) \; = \; f_K\,g \; ,
\label{gtrk}
\ee
where $f_K/f_\pi\approx1.22$ from data \cite{PDG04}. First dividing
(\ref{gtrk}) by (\ref{gtr}), the meson-quark coupling $g$ divides out:
\be
\frac{m_s+\hat{m}}{2\hat{m}} \; = \; \frac{f_K}{f_\pi} \; \approx \; 1.22
\;\;\; \Longrightarrow \;\;\; \frac{m_s}{\hat{m}} \; \approx \; 1.44 \; ,
\label{msdmh}
\ee
whereupon we deduce that
\be
m_s \; \approx \; 1.44\,\hat{m} \; = \; 486.0\;\mbox{MeV} \; .
\label{ms}
\ee
Stated alternatively,
\be
\frac{m_s+\hat{m}}{2} \; \approx \; f_K\,\frac{2\pi}{\sqrt{3}} \; \approx \;
411.6\;\mbox{MeV} \;\;\; \Longrightarrow \;\;\; m_s \; \approx \;
485.7\;\mbox{MeV} \; ,
\label{msgtr}
\ee
where we have made explicit use of the value $g=2\pi/\sqrt{3}$.

It is worth recalling that the almost pure $\bar{s}s$ vector $\phi$(1020) 
mass is about twice $m_s$, viz.\
\be
m_s \; \sim \; \frac{1}{2}\,m_{\phi(1020)} \; = \; 510\;\mbox{MeV} \; ,
\label{msphi}
\ee
as originally stated in \refc{BLP64}. (This works less well for lighter 
quarks, since $m_\rho/2 \approx$ 388 MeV is about 10\% larger than our
earlier estimate of $\hat{m}.$) 

To conclude this section, we tabulate here the predicted and measured
\cite{PDG04} magnetic moments of several ground-state baryons, as obtained
in the simple constituent quark model, with the above-derived quark masses
$m_u=335.5$ MeV, $m_d=339.5$ MeV, and $m_s=486.0$ MeV. From
Table~\ref{moments}, we see
\begin{table}[ht]
\caption{Baryon magnetic moments in constituent quark model vs.\ experiment
(in nuclear magnetons $e/2m_p$). Note that $\Sigma^0\Lambda$ stands for the
transition magnetic moment $|\mu_{\Sigma\Lambda}|$.}
\begin{indented}\item[]
\begin{tabular}{@{}ccccc}
\br
Baryon & $q_1q_2q_3$ & $9\,\mu_{\mbox{\scriptsize theor.}}/
m_{\mbox{\scriptsize proton}}$ & $\mu_{\mbox{\scriptsize theor.}}$ &
$\mu_{\mbox{\scriptsize exp.}}$ \\
\mr
$p$  &   $uud$  & $8/m_u+1/m_d$ &  2.793  &  $2.792847351\pm28\times10^{-9}$ \\
$n$  &   $udd$  & $-4/m_d-2/m_u$ & -1.850 &  $-1.9130427\pm5\times10^{-7}$ \\
$\Lambda$&$uds$ & $-3/m_s$       & -0.644 &  $-0.613\pm0.004$ \\
$\Sigma^+$&$uus$& $8/m_u+1/m_s$  & 2.700  &  $2.458\pm0.010$ \\
$\Sigma^-$&$dds$& $-4/m_d+1/m_s$ & -1.014 &  $-1.160\pm0.025$ \\
$\Sigma^0\Lambda$&$uds$&$\sqrt{3}\,(2/m_u+1/m_d)$& 1.608 & $1.61\pm0.08$ \\
$\Xi^0$ & $uss$  & $-4/m_s-2/m_u$& -1.480 &  $-1.250\pm0.014$ \\
$\Xi^-$ & $dss$  & $-4/m_s+1/m_d$& -0.551 &  $-0.6507\pm0.0025$ \\
$\Omega^-$ & $sss$  & $-9/m_s$   & -1.931 &  $-2.02\pm0.05$ \\
\br
\end{tabular}
\end{indented}
\label{moments}
\end{table}
that this naive picture works surprisingly well. Of course, we are aware 
of more sophisticated approaches to compute baryon magnetic moments (see e.g.\
\cite{IK80} and references therein). However, the reasonable overall
agreement in the table indicates that our derived constituent quark masses are
close to the ideal ones.

\section{Medium-heavy quark masses}
In the case of the charm quark, we could proceed as above, and try to determine
the constituent mass from the experimental decay constants for the $D^+$ and
$D_s^+$ mesons, using GTRs. Unfortunately, these constants have large or even
huge error bars, depending on which of the few experiments one picks. Moreover,
in the case of the $B$ mesons, no such constants have been measured so far.
Besides, it would be rather naive to think that one can simply generalize the
GTRs that work so nicely for light quarks straighforwardly to $SU(4)$ and
$SU(5)$. A fancy way to include mass and energy-scale corrections was
suggested very recently \cite{K05a}. However, for the purpose of the present
paper, we rather prefer to proceed as in (\ref{msphi}). Thus, we estimate the
loosely-bound charm and bottom masses as one-half the vector 
$\bar{c}c$ $J/\Psi(1S)$ and $\bar{b}b$ $\Upsilon(1S)$ masses, respectively, or
\bea
m_c\;\sim\;\frac{3096.92}{2}\;\mbox{MeV}\;\approx\;1550\;\mbox{MeV} \; , 
\label{mjc} \\[1mm]
m_b\;\sim\;\frac{9460.30}{2}\;\mbox{MeV}\;\approx\;4730\;\mbox{MeV} \; .
\label{mub}
\eea

As general confirmation that we are in the right ballpark (to within about
10\%), we turn to the medium-heavy, $D$ and $B$ pseudoscalar-meson mass
differences; we have the collection of values
\bea
D^+(\bar{d}c;1869.4) & - & D^0(\bar{u}c;1864.6) \; \approx \;
4.8\:\mbox{MeV} \; \sim \; m_d \: - \: m_u \; , \label{dpdz} \\
D^+_s(\bar{s}c;1968.3) & - & D^+(\bar{d}c;1869.4) \; \approx \;
98.9\:\mbox{MeV} \; \sim \; m_s \: - \: m_d \; , \label{dspdp} \\
B^0_s(\bar{b}s;5369.6) \; & - & B^0(\bar{b}d;5279.4) \; \approx \;
90.2\:\mbox{MeV} \; \sim \; m_s \: - \: m_d \; , \label{bszbz} \\
D^+_c(\bar{b}c;6400) \;\;\: & - & B^+(\bar{b}u;5279.0) \; \approx \;
1121\:\mbox{MeV} \; \sim \; m_c \: - \: m_u \; , \label{bcpbp} \\
B^0(\bar{b}d;5279.4) \; & - & D^-(\bar{c}d;1869.4) \; \approx \;
3410\:\mbox{MeV} \; \sim \; m_b \: - \: m_c \; .  \label{bzdm}
\eea

The medium-heavy baryons tell much the same story, since the data \cite{PDG04}
give
\bea
\Xi^0_c(dsc;2471.8) \; & - & \Xi^+_c(usc;2466.3)  = 5.5\pm1.8\:\mbox{MeV} 
\sim m_d-m_u \; \label{xzxp} \\ 
\Xi^+_c(usc;2466.3) & - & \Lambda^+_c(udc;2284.9)  \approx \;\;\,
181.4\:\mbox{MeV} \;\;\, \sim m_s-m_d \; \label{xplp} \\ 
\Omega^0_c(ssc;2697.5)\;&-&\Sigma^0_c(ddc;2452.2)  \approx \;\;\;
245.3\:\mbox{MeV} \;\;\; \sim 2(m_s-m_d) \; \label{ozsz} \\
\Xi^+_{cc}(ccd;3518.7) \; & - & \Sigma^0_c(ddc;2452.2) \; \approx \; 
1066.5\:\mbox{MeV} \;\; \sim m_c \: - \: m_d \; \label{xpsm} \\
\Lambda^0_b(udb;5624) \; & - & \Lambda^+_c(udc;2284.9) \; \approx \; 
3339.1\:\mbox{MeV} \;\; \sim m_b \: - \: m_c \; . \label{lblc} 
\eea
We see that the constituent $c$ and $b$ quark masses deduced from
Eqs.~(\ref{bcpbp},\ref{bzdm},\ref{xpsm},\ref{lblc}) are indeed roughly
compatible ($\pm\,$5--10\%) with our $1^{--}$ charmonium and bottomonium
estimates in Eqs.~(\ref{mjc}) and (\ref{mub}), respectively.
Unfortunately, we cannot test the idea on charm-bottom or double-bottom
baryons, as they have not yet been observed. The results for $m_s-m_d$
are less nice, as the pseudoscalar-meson estimates
(\ref{dspdp},\ref{bszbz}) yield $\sim95$ MeV, while the baryons in
Eqs.~(\ref{xplp},\ref{ozsz}) give on average $\approx142$ MeV. 
Nonetheless, the latter value is close to $(486.0-339.5)$ MeV $=146.5$ MeV
from Eqs.~(\ref{mud},\ref{ms}). Incidentally, not only do the
$\bar{q}q$ pseudoscalar/vector meson and $qqq$ baryon resonances follow this
same $u,d,s,c,b$ quark-mass pattern, but so do the $\bar{q}q$ scalars (see
\cite{SKRB03}, second paper).

\section{Very heavy top quark and the standard model}
Since the observed \cite{PDG04} $m_t=174.3\pm5.1$ GeV is so much heavier than
$m_u$, $m_d$, $m_s$, $m_c$, $m_b$, present approaches prefer linking the
heavy $m_t$ with the heavy $W$ and $Z$ bosons, in the context of the
electroweak standard model (EWSM \cite{W67}). The basic formulae are
\bea
\frac{g^2_w}{8M^2_W} \; = \; \frac{G_F}{\sqrt{2}} \; , \label{gw} \\
f_w \; = \; [\sqrt{2}\,G_F]^{-\frac{1}{2}} \;\approx\; 246.2\;\mbox{GeV}\;,
\label{fw} \\
\sin^2\!\Theta_w \; = \; \left(\frac{\bar{e}}{g_w}\right)^2 \; \approx \;
0.2284 \; \sim \; \left(1-\frac{M^2_W}{M^2_Z}\right) ,
\label{stw}
\eea
implying $g_w=0.65326$, for $G_F=11.6637(1)\times10^{-6}$ GeV$^{-2}$ and
observed $M_W=80.425(38)$ GeV, with $\bar{e}^2/4\pi\approx1/128.91$ at the
scale $M_Z=91.1876(21)$ GeV.

This is indeed a beautiful theory for the very heavy $W$ and $Z$ bosons,
wherein the $W$ and the $Z$ are treated as elementary particles;
the left-handed fermion fields transform as $SU(2)$ doublets, but the
right-handed fields as $U(1)$ singlets. What makes the theory doubly impressive
is the fact that one can compute radiative corrections reliably, because
it is renormalizable. Although the EWSM predictions are all
very accurate, they are saddled with 19 arbitrary parameters (see
\cite{PDG04}, page 160  and section VII). Furthermore, the quark masses
are ``current'' masses and do not incorporate dynamical contributions.

Instead, we attempt to reduce these 19 parameters by treating the heavy
$W$ and $Z$ bosons as resonances, like one does in strong interactions (SI),
using e.g.\ vector-meson dominance (VMD) \cite{S60} concepts and KSRF 
\cite{KS66} identities, not usually considered in the EWSM. For example, 
one knows that the $\rho$ meson approximately obeys the KSRF relation
\be
m_\rho \; = \; \sqrt{2}\,(g_{\rho\pi\pi}\,g_\rho)^{\frac{1}{2}}\,f_\pi \; ,
\label{ksrf}
\ee
where $g_{\rho\pi\pi} \simeq 5.95$ comes from the $\rho$ decay width,
\be
\Gamma_{\rho\pi\pi} \; = \; \frac{g^2_{\rho\pi\pi}}{6\pi m^2_\rho}\:q^3
\; \approx \; (150.3\pm1.6)\;\mbox{MeV} \; ,
\label{grpp}
\ee
and
$g_\rho \simeq 4.96$ is determined by the much smaller decay width,
\be
\Gamma_{\rho\,e^+\!e^-} \; = \; \frac{e^4m_\rho}{12\pi g^2_\rho} \; \approx \;
(7.02\pm0.11)\;\mbox{keV} \; .
\label{gree}
\ee
(In passing, note that the chiral QLL$\sigma$M predicts
\cite{BRS98} $g_{\rho\pi\pi}/g_\rho=6/5$, in excellent agreement with the
above data ratio.) 

The weak-interaction analogue of the KSRF relation (\ref{ksrf}) is obtained
by the substitutions
\be
m_\rho\to M_W \;\;\; , \;\;\; \sqrt{g_{\rho\pi\pi} g_\rho}\to\frac{g_w}{2}
\;\;\; , \;\;\; \sqrt{2}\,f_\pi\to f_w \; ,
\label{ksrfw}
\ee
where the weak coupling simulates $g_\rho\tau^+/2$,
and the charged $W$ requires a $\sqrt{2}$ in the weak (VEV) decay constant.
Indeed the weak KSRF relation 
\be
M_W \; = \; \frac{1}{2}\,g_w\,f_w \; ,
\label{mw}
\ee
corresponds precisely to the famous EWSM relation \cite{W67}. Other physicists 
have also searched for the relation between the EWSM and high-energy resonances
\cite{HS78}.

Extending this VMD scheme to the heavy $Z$ boson, the analogue of the
$\rho^0\to e^+\!e^-$ rate in (\ref{gree}) above determines the coupling constant
$g_Z$ of the $Z$ boson to electrons: 
\be
\Gamma_{Ze^+\!e^-} \; = \; \frac{e^2\bar{e}^2M_{Z}}{12\pi g^2_{Z}}
\; = (83.913\pm0.126)\;\mbox{MeV} \; ,
\label{gzee}
\ee
Inserting $e^2/4\pi=1/137.036$ and $\bar{e}^2/4\pi=1/128.91$ at the $Z$ mass,
we arrive at
\be
|g_Z| \; \approx \; 0.50761 \; ,
\label{gz}
\ee
Now the tree-level vector and axial-vector couplings of $Z$ to the
leptons get modified, from $g^e_V = -1/2 + 2 \sin^2\theta_w, g^e_A = -1/2$
to
\be
g^e_A \; = \; -0.50123(26) \;\;\; , \;\;\; g^e_V \; = \; -0.03783(41) \; ,
\label{geavexp}
\ee
by radiative corrections. Nevertheless, $Z$ remains largely axial (since 
\cite{PDG04} $\sin^2\theta_w = 0.23120(15)$), and the difference
$V\!-\!A$ coupling is
\be
g^e_{V\!-\!A} \; = \; 0.4633(34) \; .
\label{gevmaexp}
\ee
This is reasonably close to the EWSM value $2\sin^2\Theta_w = 0.462$ and is
also supported by the ratio of (40) to the conventional EWSM rate for $Z$,
namely
\be
\Gamma_{Ze^+e^-}=\left(\frac{g_w}{4}\right)^2 \frac{M_Z^3}{12\pi M_W^2}
 = 82.936 {\rm MeV},
\ee
which yields the alternative expression (compatible with the data above)
\be
 \sin^2\Theta_w = 1 - (g_wg_Z/4e\bar{e})^2 = 0.23118.
\ee

More interestingly, we may estimate the very heavy top quark mass $m_t$ 
via a GTR as we did for the lighter quarks. Here we have to be careful to
take account of an EWSM factor of $2\sqrt{2}$ and the (V-A) VMD coupling
$g_Z/2$. In this way we get
\be
m_t \; = \; 2\sqrt{2}\,f_w\,\frac{|g_Z|}{2} \; = \; \sqrt{2}\,
(246.2\:\mbox{GeV})\,(0.50761) \; \approx \; 176.7\;\mbox{GeV} \; , 
\label{mt}
\ee
compatible with data \cite{PDG04} at $174.3\pm5.1$ GeV.

Lastly, we examine the scalar Higgs-boson mass in the spirit of B.~W.~Lee's
null tadpole sum for the $SU(2)$ \lsm\ \cite{L72}, characterizing the true
vacuum (as obtained in  \cite{DS95,SKRB03})
\be
m^2_\sigma \; = \; 4\hat{m}^2 + m^2_\pi \; .
\label{msigma}
\ee
For the EW model the analogue of this relation is the vanishing expectation
value of the charged Higgs components; this constraint produces
\be
m^2_H \; = \; 4m^2_t-2M^2_W-M^2_Z \; \approx \; (316.7\:\mbox{GeV})^2 \; ,
\label{mhiggs}
\ee
as originally found in \cite{V81}. A somewhat smaller Higgs mass, about
216 GeV, results from a recent renomalization-group resummation of all
leading-logarithm contributions \cite{EMMS03}. Of course, we are well aware that
most physicists favour a much smaller Higgs mass, of the order of 100 GeV.
For instance, the 2004 PDG review \em ``Electroweak model and constraints on
new physics'' \em \/claims (\cite{PDG04}, page 122) \em ``The data
indicate a preference for a small Higgs mass'', \em arriving at a central
global-fit value of $M_H=113^{+56}_{-40}$ GeV. However, the detailed analyses
in \cite{C95} showed that such ``predictions'' should be taken with a
great deal of caution, leading to the conclusion that even Higgs masses in the
range 500--1000 GeV cannot be excluded on the basis of LEP data.

\section{Mixing angles and conclusion}
The 19 parameters of the EWSM (not extended to massive neutrinos) are: \\[1mm]
(a) 9 fermion masses, i.e., the six $u$, $d$, $s$, $c$, $b$, $t$ quark and the 
three $e$, $\mu$, $\tau$ lepton masses.  \\[1mm]
(b) 
3 gauge couplings, namely $\alpha_s=g^2/4\pi\approx\pi/3$, for
$g=2\pi/\sqrt{3}$ \cite{ES84,DS95}, and $g_w$, $g'_w$, with the derived
electromagnetic coupling $e=g_w g'_w/\sqrt{g^2_w+g'^2_w}$. \\[1mm]
(c) 
3 vacuum or mass scales, i.e., $v=\left<0|\phi_H|0\right> = f_w$, $\,m_H$
(our (\ref{mhiggs})), and $\Theta_{\mbox{\scriptsize QCD}}=0$ (see the $U_A(1)$
problem \cite{KKS00}). \\[1mm]
(d) 
3 quark CKM mixing angles and 1 CPV phase angle $\delta$. The data (see 
\cite{PDG04}, page 130) give:
\bea
V_{ud} & \sim & V_{cs} \sim \cos\Theta_c \sim 0.974\;\;\;\Rightarrow\;\Theta_c
\sim 13.1^\circ \; , \label{vudcs} \\
V_{us} & \sim & V_{cd} \sim \sin\Theta_c \sim 0.224\;\;\;\Rightarrow\;\Theta_c
\sim 12.9^\circ \; , \label{vuscd} \\
V_{cb} & \sim & V_{ts}\sim\sin\Theta_2\sim0.041\;\;\;\Rightarrow\;\Theta_2
\sim 2.35^\circ \; , \label{vcbts} \\
       &      & V_{tb} \sim \cos\Theta_2 \sim 0.9991 \; \Rightarrow\;\Theta_2
\sim 2.43^\circ \; , \label{vtb} \\
V_{td} & \sim & V_{ub} \sim \sin\Theta_3 \sim 0.0066 \; \Rightarrow\;\Theta_3
\sim 0.38^\circ \; . \label{vtdub} 
\eea
so we may conclude that $\Theta_c\sim13^\circ$, $\Theta_2\sim2.39^\circ$,
$\Theta_3\sim0.38^\circ$. Lastly, the angle $\delta$, has been measured as
\cite{PDG04} $\delta=(3.27\pm0.12)\times10^{-3}$. Note, too, that
$\Theta_c/\Theta_2:\Theta_2/\Theta_3\sim6$.

Given the consistent pattern of the 6 quark masses in Secs.~3--6, we try
to make use of them in the manner of Fritzsch \cite{F77}, but
adopting instead {\em constituent} quark masses. Using our values for $m_q$,
this approach predicts \cite{PS85} $\phi_{sd}=22.9^\circ$, and approximately
\bea
\sin2\phi_{cu} \; \approx \; \sqrt{\frac{m_s-m_d}{m_c-m_u}} \; \approx \;
0.347 \;\;\; \mbox{or} \;\;\; \phi_{cu} \; \approx \; 10.2^\circ \; , 
\label{phicu} \\[1mm]
\Theta_c \; = \; \phi_{sd} - \phi_{cu} \; \approx \; 22.9^\circ-10.2^\circ
\; = \; 12.7^\circ \; , \label{thetac} \\[1mm]
\sin2\Theta_2 \; \approx \; \sqrt{\frac{m_c-m_u}{m_t-m_c}} \; \approx \;
0.083 \;\;\; \mbox{or} \;\;\; \Theta_2 \; \approx \; 2.4^\circ \; , 
\label{theta2} \\
\Theta_3 \; \sim \; \,\Theta_2/6 \; \sim \; 0.4^\circ \; ,
\label{theta3} 
\eea
Thus we see that the predicted CKM angles $\Theta_c$ and $\Theta_2$
are reasonably near the observed ones in
(\ref{vudcs}--\ref{vtb}). Furthermore, the CPV phase angle $\delta$
can be estimated as \cite{CS96}
\be
\delta \; = \; \frac{\alpha}{\pi}\,\ln\left(1+\frac{\Lambda^2}{m^2_t}\right)
\; = \; 3.34\times10^{-3}\; ,
\label{delta}
\ee
taking for $m_t$ our predicted value of about 177 GeV (see (\ref{mt})), and
for the UV cutoff $\Lambda$ the Higgs mass $m_H\approx 317$ GeV, as derived in 
(\ref{mhiggs}). This value for $\delta$ is compatible with data
\cite{PDG04} at $(3.27\pm0.12)\times10^{-3}$. 

In conclusion, in the present paper we have employed constituent quark masses,
instead of the usual current masses, to reduce part of the arbitrariness of
the 19 parameters in the SM. Our nonstrange constituent quark mass follows
from the QLLSM GTR $g f_\pi\!=\!\hat{m}$. A further justification for
using constituent quark masses is S.~Weinberg's mended-chiral-symmetry paper
\cite{W90a}, which predicted the sigma width as 9/2 times the rho width,
resulting in the value $4.5\times(150.3\pm1.6)$ MeV $=(676.4\pm7.2)$ MeV,
astonishingly near the QLLSM prediction for the sigma mass in terms of
constituent quark masses: $m_\sigma=2\times337.5$ MeV $=675$ MeV. This
(near) equality of the sigma mass and width is crucial to obtain the
correct amplitude magnitude for the $\Delta I\!=\!1/2$ \em weak \em
$K_S\to2\pi$ decay, via a sigma-pole graph (see e.g.\ Ref.\ \cite{BKRS02}).
Then, in his immediately following paper \cite{W90b}, Weinberg stated in the
abstract: \em ``An explanation is offered why quarks in the constituent quark
model should be treated as particles with axial coupling $g_A\!=\!1$ and no
anomalous magnetic moment.'' \em Well, note that our constituent-quark GTRs
in Eqs.~(\ref{gtr},\ref{gtrk}) do indeed have $g_A\!=\!1$ at this quark level.
Summarizing, constituent quark masses always make chiral contact with
data, while current quark masses are more problematic.

Wrapping up, the analysis of this paper allows us to fix 13 of the 19 
parameters from the experimental data \cite{PDG04} and also provides a link
with the present EWSM. Moreover, employing once again ($V\!-\!A$)
currents, we can derive \cite{SM58,FGM58} two additional relations among $G_F$,
$m_\mu$, and $m_e$, thus further reducing the arbitrariness of the remaining
6 parameters of the Standard Model.\\[1mm]

\ack
One of us (MDS) is indebted to V.~Elias for prior discussions and
collaboration. We also thank F.~Kleefeld for very useful comments. This work
was partly supported by the
{\it Funda\c{c}\~{a}o para a Ci\^{e}ncia e a Tecnologia}
\/of the {\it Minist\'{e}rio da Ci\^{e}ncia, Tec\-nologia e 
Ensino Superior} \/of Portugal,
under contracts POCI/FP/63437/2005 and POCI/FP/63907/2005.

\end{document}